\documentclass[twocolumn,aps,prb,superscriptaddress,amsmath,amssymb,showpacs,floatfix,preprintnumbers]{revtex4-2}
\usepackage[latin9,utf8]{inputenc}
\usepackage{amsmath,amssymb,amsfonts,dsfont,mathrsfs,amsthm}
\usepackage{graphicx}
\usepackage{centernot}
\usepackage{xcolor}
\usepackage{comment}
\usepackage{hyperref}
\hypersetup{linktocpage,colorlinks=true,urlcolor=blue,linkcolor=blue,citecolor= blue}
\usepackage{feynmf}
\usepackage{siunitx}
\usepackage{array}
\usepackage{tikz}
\usepackage{braket}
\usepackage{chemformula}

\begin{document}

\title{Berezinskii-Kosterlitz-Thouless transition with enhanced phase stiffness\\ in $d$-wave strongly coupled two-dimensional superconductors}
\author{Sathish Kumar Paramasivam}
\address{COMMIT, Department of Physics, University of Antwerp, Groenenborgerlaan 171, 2020 Antwerp, Belgium.}
\address{CQM group, School of Science and Technology, Physics Division, University of Camerino, 62032 Camerino (MC), Italy.}

\author{Andrea Perali}
\email{andrea.perali@unicam.it}
\address{CQM group, School of Pharmacy, Physics Unit, University of Camerino, 62032 Camerino (MC), Italy.}

\author{Milorad V. Milo\v{s}evi{\'c}}
\email{milorad.milosevic@uantwerpen.be}
\address{COMMIT, Department of Physics, University of Antwerp, Groenenborgerlaan 171, 2020 Antwerp, Belgium.}

\begin{abstract}
We reveal the key role of the $d$-wave symmetry of the superconducting gap in strongly coupled two-dimensional (2D) superconductors in determining the properties of the 
Berezinskii-Kosterlitz-Thouless (BKT) transition, associated with a sizable amplification of the  phase stiffness with respect to nodeless-gap superconductors. The enhanced stiffness originates from extended regions of vanishing gap around the nodal lines of the Brillouin zone (BZ). Our study, based on mean-field and BKT theory, presents a comparative analysis of $s$-wave and $d$-wave scenarios, highlighting the features of the latter that boost the stiffness and the BKT transition temperature (T$_{\text{BKT}}$). The comparison centers on two quantities: the mean-field critical temperature and the maximum superconducting gap related to the pairing strengths. We present a phase diagram that captures the scaling of T$_{\text{BKT}}$ with respect to the mean-field critical temperature across the BCS–BE crossover and the evolution of the pseudogap. The zero-temperature phase stiffness intensity map over the BZ is also presented, with a distinctly two-component structure consisting of low- and high-stiffness regions, whose extent depends on microscopic system parameters. These results identify the nodal gap structure of strongly coupled 2D superconductors as a likely enabler for enhanced stiffness and T$_{\text{BKT}}$ compared to their $s$-wave counterparts.
\end{abstract}
\maketitle
\section{Introduction}
Motivations to investigate the Berezinskii-Kosterlitz-Thouless (BKT) transition in 2D $d$-wave superconductors are manifold. A key motivation stems from the recent fabrication of monolayers of Bi$_2$Sr$_2$CaCu$_2$O$_{8+x}$ (BSCCO) $d$-wave cuprate superconductors \cite{mono1, mono2, spatially, D.jiang_2014} and their exploitation in quantum technology, as superconducting qubits \cite{d_wave_qubits, d-wave} based on twisted 2D van der Waals (vdW) heterostructures of few-layer cuprates \cite{Uri}. BSCCO cuprate has the d$_{x^{2}-y^{2}}$ pairing symmetry \cite{sop} and displays superconductivity even when reduced to a half-unit cell layer. Due to its intrinsic 2D nature, superconductivity in these systems can occur below the BKT transition temperature. The BKT transition in 2D cuprates is probed via the non-linear exponent of the current-voltage (I-V) characteristics, which provides a clear signature of the BKT transition. Namely, in the power-law dependence V $\propto$ I$^\alpha$, the exponent $\alpha$ abruptly changes from 1 to 3 at the BKT transition temperature T$_{\text{BKT}}$ \cite{G_venditti,M.Sharma}. The ultra thin BSCCO films grown on SrTiO$_{3}$ substrate exhibit T$_{\text{BKT}}\approx89$~K \cite{zhang_2023}. Other experiments on cuprate ultra-thin films of different thickness report T$_{\text{BKT}}$ from 85 to 89.1~K \cite{mono2,s_wang_2024}. Likewise, a BKT transition of a two unit-cell thick Bi$_2$Sr$_2$Ca$_2$Cu$_3$O$_{10+\delta}$ (Bi-2223) microbridge has been reported at approximately 80~K \cite{A.B.Yu}. Interestingly, the twisted interfaces between cuprate single-crystals have been proposed as a tunable platform for Josephson junctions with controllable transversal pairing, a desirable feature for the superconducting qubits. Recent experimental realizations have been discussed in Refs. \cite{Uri, 2d_Poccia, N.Poccia}, where the $d$-wave symmetry of the (quasi) order parameter plays a fundamental role in the two-Cooper pair tunneling, while the large gap values protect the qubits against decoherence. 

Beyond cuprates, the bilayer nickelate La$_3$Ni$_2$O$_7$ \cite{Z.Fan_2024, S.Ryee_2025, D_K_Singh, zhang_2024} also exhibits a quasi-2D crystal structure and is proposed to host the $d$-wave gap symmetry. Furthermore, the heavy fermion superconductors CeXIn$_{5}$ (X = Co, Ir, Rh) \cite{R.Movshovich_2001,k.izawa, G.sey,Truncik, h.hegger} and the layered organic superconductors $\kappa$-(BEDT-TTF)$_2$X (X = Cu[N(CN)$_2$]Br, Cu(SCN)$_2$) \cite{Organic_SC_0,Organic_SC_1,Organic_SC_2,Organic_SC_3,Organic_SC_4}  are found to exhibit quasi-two dimensional behavior and $d$-wave order parameter, with unconventional pairing mechanism driven by spin or charge fluctuation–mediated interaction. Therefore, any of the aforementioned quasi-2D material classes may present a platform for investigating BKT physics considering different symmetries of the order parameter. Yet another intriguing $d$-wave superconducting phase has been discussed in graphene monolayers with Van Hove singularity (VHs) \cite{R.Thomale,Chubukov}, and its BKT transition has been explored as a function of applied strain \cite{BKT_graphene}. Ref.~\cite{Biplab} presented a theoretical study of the BKT transition properties in quasi-2D Bi2212 cuprates. They confirmed that the $d$-wave order parameter is in the best agreement with the experimental data, after examining the effects of various gap symmetries across different doping levels.

\begin{figure}[t]
\centering
\includegraphics[width=0.75\linewidth]{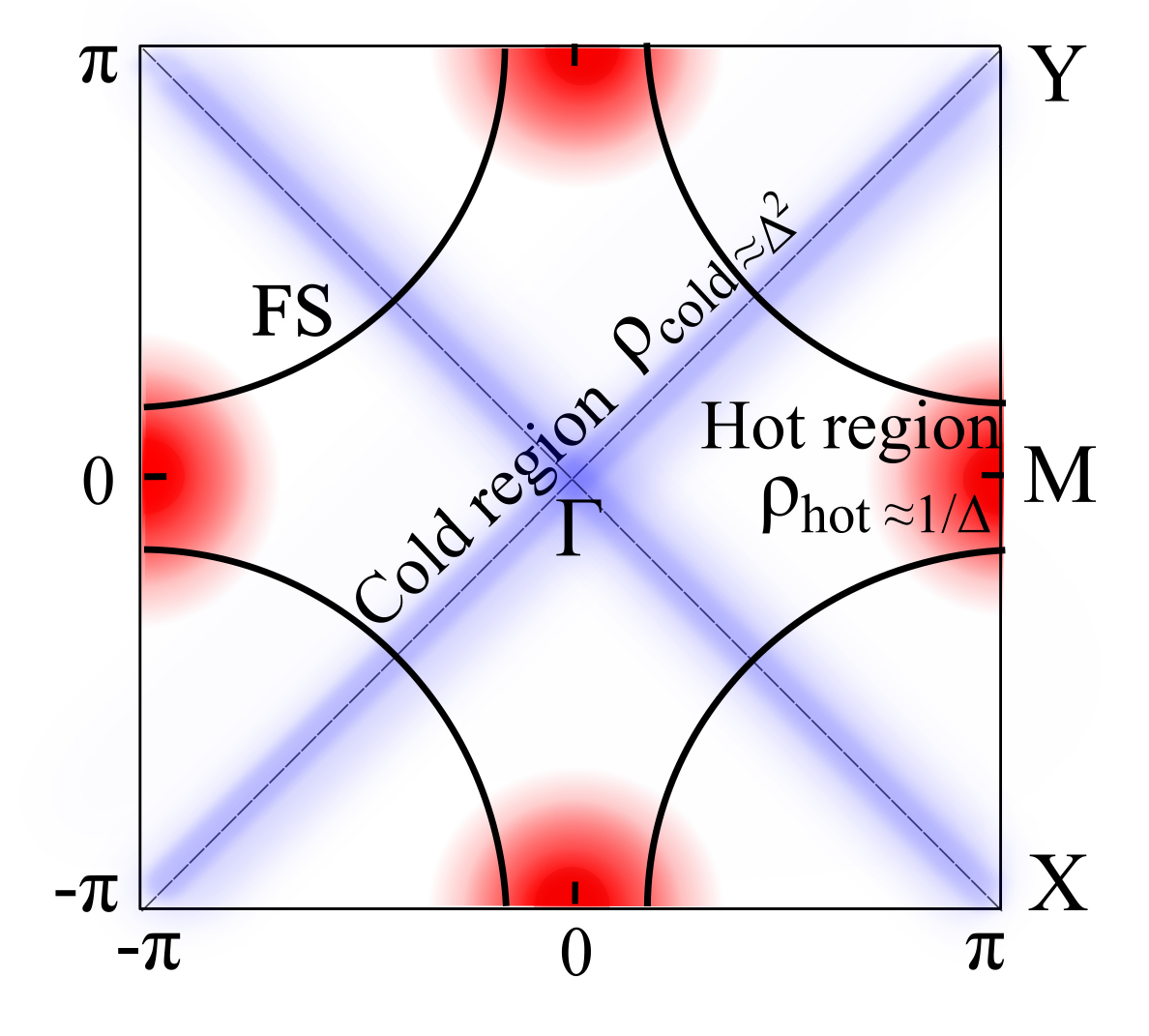}
 \caption{Brillouin zone and schematic Fermi surface of a $d$-wave cuprate superconductor close to optimal doping, with hot region (red patches) and cold region (blue patches).} \label{fig.1}
\end{figure}

Additional interest in exploring $d$-wave symmetry within the context of BKT phase transition in 2D superconductors around the VHs, particularly in the strong-pairing regime, originates from the insights provided by the two-patch model to effectively describe the wave-vector dependent properties of the excitation spectrum and segmented Fermi surfaces of cuprates, as discussed in Ref.~\cite{perali_dwave,multi-patch,Two-gap,Two-gap_stripe}. This investigation is closely tied to understanding the phenomenology of cuprate superconductors~\cite{scalapino, sop,shen_1993,H.ding_1996}. In the two-patch model, both the BZ and the Fermi surface (FS) are partitioned into two regions. The hot patches correspond to the areas around the M points in the BZ shown in Fig.~\ref{fig.1}. The electronic band dispersion exhibits saddle points located at these M points, which lead to pronounced VHs in the density of states (DOS)~\cite{saddle_vhs, gap_vhs_1995}. The electronic states near the M points are nearly incoherent and have a reduced Fermi velocity. These hot states lead to formation of local Cooper pairs behaving as Bose-Einstein (BE)-like molecules. The cold patches emerge around the nodal lines of the BZ along the $\Gamma$Y(X) diagonal shown in Fig.~\ref{fig.1}. The quasiparticles around the nodes are coherent with high Fermi velocities. These states correspond to cold states, forming extended Cooper pairs with BCS-like character. The strongly anisotropic pairing interaction generates diverse pairing strengths across different arcs of the segmented Fermi surface, resulting in partial superconducting (quasi) condensates coexisting within the BZ~\cite{vishik_2010}. 

Using a  mean-field analysis with BKT theory, we compare the effects of s-wave and d-wave pairing symmetries in strongly coupled 2D superconductors. In particular, we analyze how these symmetries influence key physical quantities such as the phase stiffness, the BKT transition temperature with respect to the maximum superconducting gap, and the mean-field critical temperature. The effective pairing interactions are assumed to be of electronic origin, such as charge or spin fluctuations coupled to high-energy phonons that span the full electronic bandwidth. This is relevant for cuprates and even more for superconductors characterized by quasi-flat bands crossing or approaching the chemical potential. We reveal that anisotropic $d$-wave gap structure plays a key role in amplifying the phase stiffness and boosting the BKT transition temperature. This is visualized through an intensity plot of the zero-temperature stiffness in the BZ, which clearly exposes the symmetry driven variation in stiffness. The phase diagram of the system reveals that in the strong-pairing BE regime the $d$-wave symmetry determines the shrinking of the pseudogap with respect to the $s$-wave case. Indeed, the effective two-component behavior induced by the $d$-wave gap leads to physics analogous to the screening of the superconducting fluctuations in multiband systems with flat and deep electronic bands contributing to the condensate~\cite{L.Salasnich_2019,sathish}. Our results can be extended beyond single-band 2D $d$-wave superconducting systems, offering a general framework applicable to multi-band and multi-gap superconductors with at least one $d$-wave partial condensate.

The manuscript is organized as follows. Section II details the physical system and the mean-field theoretical approach for evaluating the superconducting state properties; a summary of the BKT theory is also provided. Section III contains our results, discussion and analysis with regard to the superconducting phase stiffness and the BKT transition temperature, systematically comparing the $d$-wave with the $s$-wave pairing symmetry. Our conclusions are summarized in Section IV.
\begin{figure}[t]
\centering
\includegraphics[width=\linewidth]{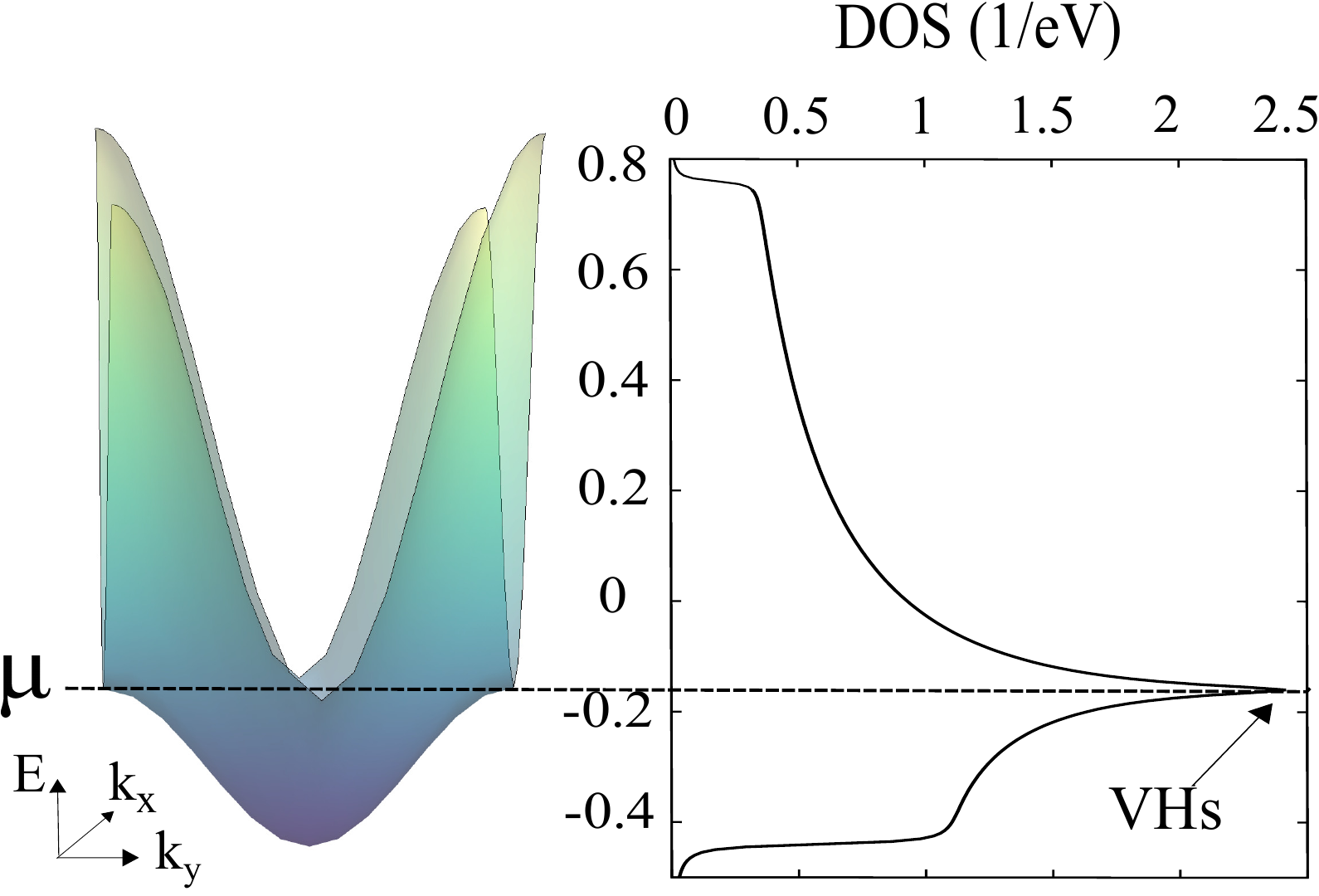}
 \caption{The tight binding energy dispersion with nearest and next-nearest neighbor hopping parameters $t = 0.15$~eV and $t^{\prime}$= 0.04~eV. The chemical potential ($\mu$) is fixed near the VHs, at -4t$^{\prime}$.}
 \label{fig.1.1}
\end{figure}

\section{Theoretical framework}
We consider a single-band 2D tight-binding model on the square lattice with energy dispersion determined by hopping up to the second-nearest neighbors:
\begin{equation}
\begin{split}
    \epsilon(k) &= -2t \left[ \cos(k_{x}a) + \cos(k_{y}a) \right] \\
    &\quad + 4\mathrm{t}^{\prime} \left[ \cos(k_{x}a) \cos(k_{y}a) \right],
\end{split}
\end{equation}
where $a$ is the lattice parameter, $t$ and t$^{\prime}$ are the nearest neighbor and next-nearest neighbor hopping parameters, respectively, with values of $t$ = 0.15~eV and t$^{\prime}$ = 0.04~eV. These parameters are appropriate to incorporate the main features of the band structure of the Bi-2212 compounds~\cite{perali_dwave, Randeria} in a simplified model, even though the application of our approach is not limited to cuprates, but also includes other 2D superconducting systems with quasi-flat bands and $d$-wave gaps. The full bandwidth is 1.20 eV and the wave-vectors belong to the first BZ $-\frac{\pi}{a} \leq k_{x,y} \leq \frac{\pi}{a}$. Fig.~\ref{fig.1.1} illustrates the tight-binding energy dispersion and the DOS. The fixed chemical potential aligns with the energy dispersion at the point where the VHs appears in the DOS. This choice is deliberate, focusing on the regime of interest that we aim to explore. In what follows, we use units $\hbar$ = $\textit{a}$ = $k_{B}$ = 1. We numerically solve the self-consistent BCS mean-field gap equation at a finite temperature, concurrently with the density equation
\begin{equation}\label{gap}
 \begin{split}
  \Delta(\textbf{k}, T) = & -\frac{1}{\Omega} \sum_{\boldsymbol{k^{'}} } \Biggr[  V(\boldsymbol{k},\boldsymbol{k^{'}})  \frac{ \tanh{ \frac{E({\boldsymbol{k^{'}}})}{2T}} \Delta( \boldsymbol{k^{'}} ) } {2E({\boldsymbol{k^{'}}})  }    \Biggr],\
\end{split}
\end{equation}
\begin{equation}\label{n_density}
\begin{split}
 n = & \frac{2}{\Omega}\sum_{\boldsymbol{k}}  \Biggr[  \frac{1}{2}\Bigl( 1-  \frac{ \xi(\boldsymbol{k})}{  E({\boldsymbol{k^{}}}) 
  }  \Bigl)f(-E({\boldsymbol{k}}))\\    
    &+\frac{1}{2} \Bigl( 1+  \frac{ \xi(\boldsymbol{k})}{  E({\boldsymbol{k^{}}}) }   \Bigl)f(E({\boldsymbol{k}})) \Biggr].
\end{split}
\end{equation}
The factor 2 in Eqn.~(\ref{n_density}) is due to spin degeneracy. E$_{\boldsymbol{k}}$=$\sqrt{\xi(\boldsymbol{k})^{2}+\Delta(\boldsymbol{k})^{2}}$ is the dispersion of single-particle excitations in the superconducting state. $\Omega$ is the area occupied by the 2D superconducting system. $\xi(\boldsymbol{k}) = \epsilon(\boldsymbol{k})-\mu$ is the dispersion relation for the electronic band with respect to the chemical potential. In the single-band case, the dimensionless electron pairing constant is defined as ${\lambda}$ = N(E$_{0}$)V$_{0}$, where N(E$_{0}$) is the DOS at the bottom of the band. $f(x) =[1 + \exp(x/T)]^{-1}$ is the Fermi-Dirac distribution function. At T = 0, the gap Eqn.~(\ref{gap}) and density Eqn.~(\ref{n_density}) can be written as
\begin{equation}\label{gap_0}
 \begin{split}
  \Delta(\textbf{k}) = & -\frac{1}{\Omega} \sum_{\boldsymbol{k^{'}} } \Biggr[  V(\boldsymbol{k},\boldsymbol{k^{'}})  \frac{  \Delta( \boldsymbol{k^{'}} ) } {2E({\boldsymbol{k^{'}}})  }    \Biggr],\
\end{split}
\end{equation}
\begin{equation}\label{n_density_0}
\begin{split}
 n = & \frac{2}{\Omega}\sum_{\boldsymbol{k}}  \Biggr[  \frac{1}{2}\Bigl( 1-  \frac{ \xi(\boldsymbol{k})}{  E({\boldsymbol{k^{}}}) 
  }  \Bigl) \Biggr].
\end{split}
\end{equation}
In 2D systems, the superconducting phase transition takes place at the BKT transition temperature T$_{\text{BKT}}$ \cite{Bkt_1, Bkt_2, Bkt_3, Bkt_4}. Below the BKT temperature, the 2D system exhibits superconductivity with quasi-long-range order: the superconducting order has a power-law decay and the order parameter can be considered as a quasi-order parameter, with superconducting state properties that can be measured locally in space \cite{Luca_Salasnich_2024}. The BKT phase transition in 2D superconductors with $s$-wave symmetry has been extensively investigated in the literature, particularly in the contexts of disorder \cite{disorder, disorder_1, disorder_2}, multiband scenarios \cite{Midei_andrea, midei_2024, sathish}, heterostructures \cite{hetrostructure, hetrostructure_1, hetrostructure_2} and in topological bandstructures \cite{Q.Gao, P_torma_1, P_torma_2}. The phenomenology becomes more complex for the $d$-wave channel due to its nodes in the gap function at specific wave-vectors of the BZ. The BKT transition temperature T$_{\text{BKT}}$ can be determined through self-consistent solutions of the Kosterlitz-Thouless condition \cite{Bkt_2}
\begin{equation}\label{4}
    T_{BKT} =  \frac{\pi}{2} \rho_{s}(\Delta(T_{BKT}),\mu(T_{BKT}),T_{BKT}),
\end{equation}
where $\rho_{s}(T)$ is the phase stiffness, $\Delta$ and $\mu$ are temperature dependent and given by the solutions of the gap Eqn.~\eqref{gap} and the density Eqn.~\eqref{n_density}. Phase stiffness at BCS mean-field level for a generic band is given by \cite{Walter,sf_benfatto}
\begin{equation} \label{rho_s}
\begin{split}
     \rho_{s}(T) & = \frac{1}{4}\int \frac{d^{2}k}{(2\pi)^{2}} \frac{\partial^{2}\epsilon(\boldsymbol{k}) }{\partial k_{\alpha}^{2}} \left[ 1 - \frac{\xi({\boldsymbol{k}}) }{E_{\boldsymbol{k}}}\tanh\left({\frac{E_{\boldsymbol{k}}}{2T}}\right) \right] \\
 &+ \frac{1}{2} \int \frac{d^{2}k}{(2\pi)^{2}} \left( \frac{\partial\epsilon(\boldsymbol{k})}{\partial k_{\alpha}}  \right)^{2}\frac{\partial f (E_{\boldsymbol{k}})}{\partial E_{\boldsymbol{k}}},
 \end{split} 
\end{equation}
where $k_{\alpha} = k_{x,y}$. Due to the tetragonal symmetry of the system, $\rho_{x,x}$ = $\rho_{y,y}$ = $\rho_{s}$.

At T = 0, we can write the above Eqn.~(\ref{rho_s}) as
\begin{equation}\label{rho_s_T=0}
     \rho_{s}(0)  = \frac{1}{4}\int \frac{d^{2}k}{(2\pi)^{2}} \frac{\partial^{2}\epsilon(\boldsymbol{k}) }{\partial k_{\alpha}^{2}} \left[ 1 - \frac{\xi({\boldsymbol{k}}) }{E_{\boldsymbol{k}}}   \right]. 
\end{equation}

The symmetry of the superconducting gap elucidates the features of the quantum mechanical wave function that determine the pairing of electrons and their coherence in a superconductor. The $s$-wave superconductivity emerges in systems with pairing due to electron-phonon coupling, while the $d$-wave superconductivity is the most energetically favorable ground state in electronic systems characterized by strong local repulsions, such as cuprates. Here, we focus on the d$_{x^{2}-y^{2}}$ symmetry to describe the important properties of strongly coupled 2D superconductors, such as phase stiffness and BKT critical temperature, through a systematic comparison between $s$-wave and $d$-wave cases.

The $s$-wave symmetry refers to a superconducting pairing symmetry where the angular part of the  wave function governing the pairing of electrons is spherically symmetric throughout the Fermi surface. In the $s$-wave case, the effective pairing interaction can be approximated by a separable potential $V(\boldsymbol{k},\boldsymbol{k^{'}})$ with an energy cut-off $\omega_{0}$, and it is given by:
\begin{equation}\label{s_wave_v}
V(\boldsymbol{k},\boldsymbol{k^{'}}) = -V_{0}\Theta(\omega_{0}-|\xi_{\boldsymbol{k}}|)\Theta(\omega_{0}-|\xi_{\boldsymbol{k^{'}}}|),
\end{equation}
where $V_{0} > 0$ is the strength of the attractive potential. In this way, the superconducting gap has the same energy cut-off and the same wave-vector dependence as the interaction:
\begin{equation}
    \Delta(\boldsymbol{k}, T) = \Delta(T) \Theta(\omega_{0}-|\xi_{\boldsymbol{k}}|).  
\end{equation}
In this article, the cut-off frequency $\omega_{0}$ will remain consistently fixed at 4$t$ throughout the $s$-wave calculations. We refer to effective pairing interactions of electronic origin and, therefore, the energy range of paired states includes the entire band dispersion (for both $s$-wave and $d$-wave cases). For the $s$-wave case, the order parameter is uniform across the entire Fermi surface. Consequently, there are no quasiparticle states below the energy gap.

The $d$-wave symmetry characterizes a type of superconducting pairing symmetry in which the angular portion of the quantum mechanical wave function undergoes a sign change with a rotation of 90 degrees \cite{sign_change}. This specific symmetry results in a four-fold pattern in the superconducting gap parameter. In the $d$-wave case, the effective pairing wave-vector-dependent interaction is given by:
\begin{equation}\label{d_gap_v}
V(\boldsymbol{k},\boldsymbol{k^{'}}) = -V_{0} [\cos(k_{x}a) - \cos(k_{y}a)] [\cos(k^{\prime}_{x}a) - \cos(k^{\prime}_{y}a)],
\end{equation}
where $V_{0} > 0$ is the strength of the attractive potential. The $d$-wave superconducting gap takes the form:
\begin{equation}\label{d_gap}
    \Delta(\boldsymbol{k}, T) = \Delta(T) [\cos(k_{x}a) - \cos(k_{y}a)].  
\end{equation}
The $d$-wave symmetry results in the formation of nodes on the Fermi surface where the gap parameter is zero. The nodal structure holds significant implications for the phase stiffness in 2D superconductors and the behavior of excited quasiparticles.
\begin{figure}[b] 
\centering
\includegraphics[width= \linewidth]{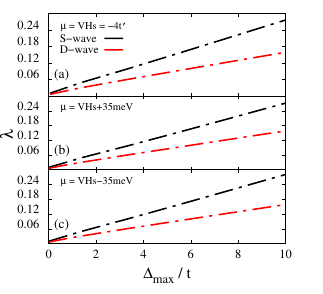}
 \caption{The pairing strength ($\lambda$) as a function of the maximum value of the superconducting gap scaled to $t$ = 0.15 eV.}\label{fig.2}
\end{figure}

We determine $s$-wave and $d$-wave superconducting gaps by solving the self-consistent BCS mean-field gap Eqn.~(\ref{gap_0}), which is coupled with the density Eqn.~(\ref{n_density_0}) for the respective potential as explained above. Here, we consider two observables to compare the $s$-wave and $d$-wave symmetries: the mean-field transition temperature (T$_{c}$) and the maximum superconducting gap. The maximum gap can be experimentally probed near-zero temperatures in the vicinity of the M points of the BZ. In 2D, the mean-field T$_{c}$ corresponds to the onset temperature of Cooper-pair formation without global phase coherence of the gap parameter (the pseudogap). This provides a useful reference for comparison with the actual critical temperature required to stabilize superconductivity, namely the BKT transition temperature, thereby offering insights into the emergence of the pseudogap.

\begin{figure}[t]
\centering
\includegraphics[width= \linewidth]{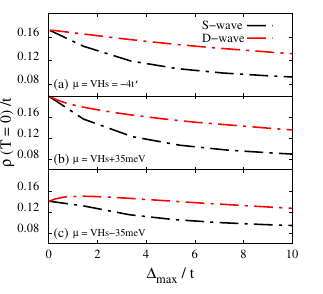}
 \caption{The phase stiffness at zero temperature as a function of the maximum superconducting gap in units of $t$ = 0.15 eV.}\label{fig.3}
\end{figure}

\section{Results and Discussion}
The pairing strengths of the $s$-wave pairing and the $d$-wave pairing should not be regarded in the same manner, because of the $k$-dependent potential associated with the $d$-wave symmetry. The pairing strengths of the $d$-wave and $s$-wave symmetry are derived from the same maximum superconducting gap $\Delta_{max}$. In Fig.~\ref{fig.2}, the pairing strength of $d$-wave and $s$-wave pairing symmetry is reported as a function of $\Delta_{max}$ in units of $t$. In Fig.~\ref{fig.2}(b, c), $\mu$ is placed above and below the VHs by 35 meV. This corresponds to the experimentally measured separation between the Fermi level and the VHs, a value associated with the optimal doping regime in the Bi2212 cuprate superconductor. The plot indicates that, having the same maximum superconducting gap, the $s$-wave corresponds to a higher pairing strength compared to the $d$-wave. 

From Fig.~\ref{fig.3}, one sees that in the vicinity of VHs the phase stiffness at zero temperature in the $d$-wave channel exceeds that of the $s$-wave pairing within the strong-pairing regime. Both phase stiffness and T$_{\text{BKT}}$ transition temperature experience significant suppression around VHs due to strong quasiparticle phase fluctuations for the $s$-wave pairing channel in the strong-pairing regime~\cite{sathish}. However, this is not the case for the $d$-wave pairing symmetry. The larger phase stiffness in the $d$-wave case is due to the co-existence of two distinct superconducting quasi-condensates in the BZ, as discussed below.

\begin{figure}[b]
\centering
\includegraphics[width=\linewidth]{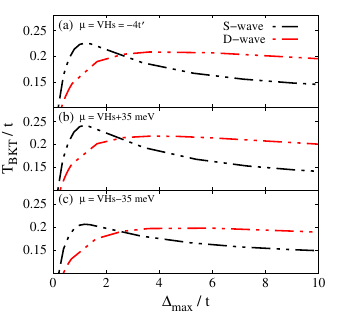}
 \caption{T$_{\text{BKT}}$ as a function of the maximum superconducting gap in units of $t$ = 0.15 eV, around VHs. (a)-(c) Chemical potential $\mu$ is fixed at VHs, VHs+35 meV, and VHs-35 meV, respectively.}
    \label{fig.4}
\end{figure}

\begin{figure}[b]
    \centering
    \includegraphics[width=\linewidth]{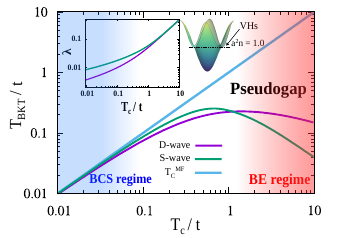}
    \caption{The BKT transition temperature T$_{\text{BKT}}$ as a function of the mean-field temperature T$_{\text{c}}$ for a normalized fixed carrier density. The figure spans the BCS-BE crossover regime and highlights the differences between $s$-wave and $d$-wave pairing symmetries. Inset shows the pairing strength as a function of the mean-field critical temperature.}
    \label{fig.5}
\end{figure}

\begin{figure}[b]
    \centering
    \includegraphics[width=\linewidth]{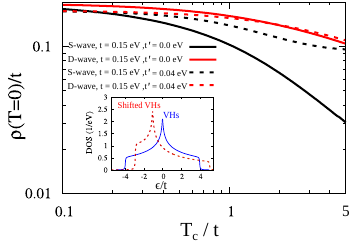}
    \caption{The dependence of phase stiffness on the mean-field critical temperature, with and without the next-nearest neighbor hopping. Inset shows the band dispersion density of states with VHs.}
    \label{fig.6}
\end{figure}

Fig.~\ref{fig.4} shows that T$_{\text{BKT}}$ for the $d$-wave symmetry is higher than that for the $s$-wave symmetry around the VHs in the strong-pairing regime. In the very weak-pairing regime, both symmetries exhibit nearly the same BKT transition temperature. As the gap increases, the $s$-wave channel becomes dominant over the $d$-wave channel. The enhanced gap scale in the strong-pairing regime is accompanied by a loss of phase coherence, leading to suppressed phase stiffness and a reduced BKT transition temperature in the $s$-wave channel. In contrast, the $d$-wave channel exhibits an enhancement under the same conditions. This enhancement in the BKT temperature is well explained by the two-patch model.

The total phase stiffness in Eqn.~(\ref{rho_s}) consists of two components, namely the diamagnetic and paramagnetic contributions. The diamagnetic response is determined by the effective mass and the total carrier density, whereas the major paramagnetic response contribution comes from the band-dependent Fermi velocity. At zero temperature, the paramagnetic response vanishes, and the total stiffness is exclusively determined by the diamagnetic response. In the case of $d$-wave pairing symmetry, the presence of nodal points leads to reduced phase fluctuations and stabilizes the quasi-condensates, resulting in a higher local phase stiffness along the nodal directions.
In contrast, the quasi-condensates near the M points exhibit stronger Cooper pairing but also enhanced phase fluctuations, which ultimately suppress the local phase stiffness in those regions. Consequently, the total phase stiffness in $d$-wave superconductors can remain significant even in the strong-pairing regime. Therefore, the coexistence of the two distinct superconducting quasi-condensates within a single-band superconducting system stabilizes the BKT transition temperature via enhancing the phase stiffness of the system by circumventing the superconducting fluctuations. In general, such coexistence of distinct quasi-condensates in a superconducting system is known to enhance the critical temperature by shrinking the pseudogap, as has been discussed in detail in Refs.~\cite{sathish,L.Salasnich_2019,T.T.Saraiva_2020,A.A.Shanenko_2022}.

\begin{figure}[b]
    \centering
    \includegraphics[width=\linewidth]{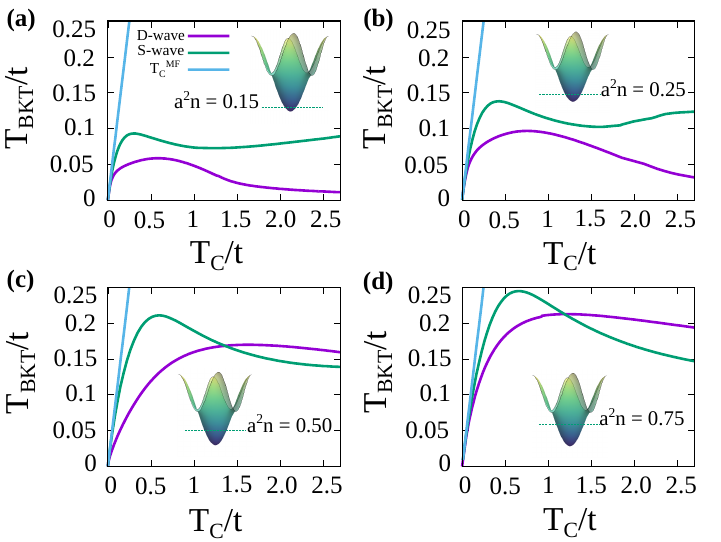}
    \caption{The BKT transition temperatures versus mean-field critical temperatures for $s$-wave and $d$-wave symmetry, with nearest-neighbor hopping. Panels (a)–(d) correspond to increasing the normalized carrier densities, represented by a$^{2}$n= 0.15, 0.25, 0.50, and 0.75, respectively, and reveal the interplay between the two symmetries across different density regimes.}
    \label{fig.7}
\end{figure}

\begin{figure*}[htbp]
\centering
\includegraphics[width = 0.75\textwidth]{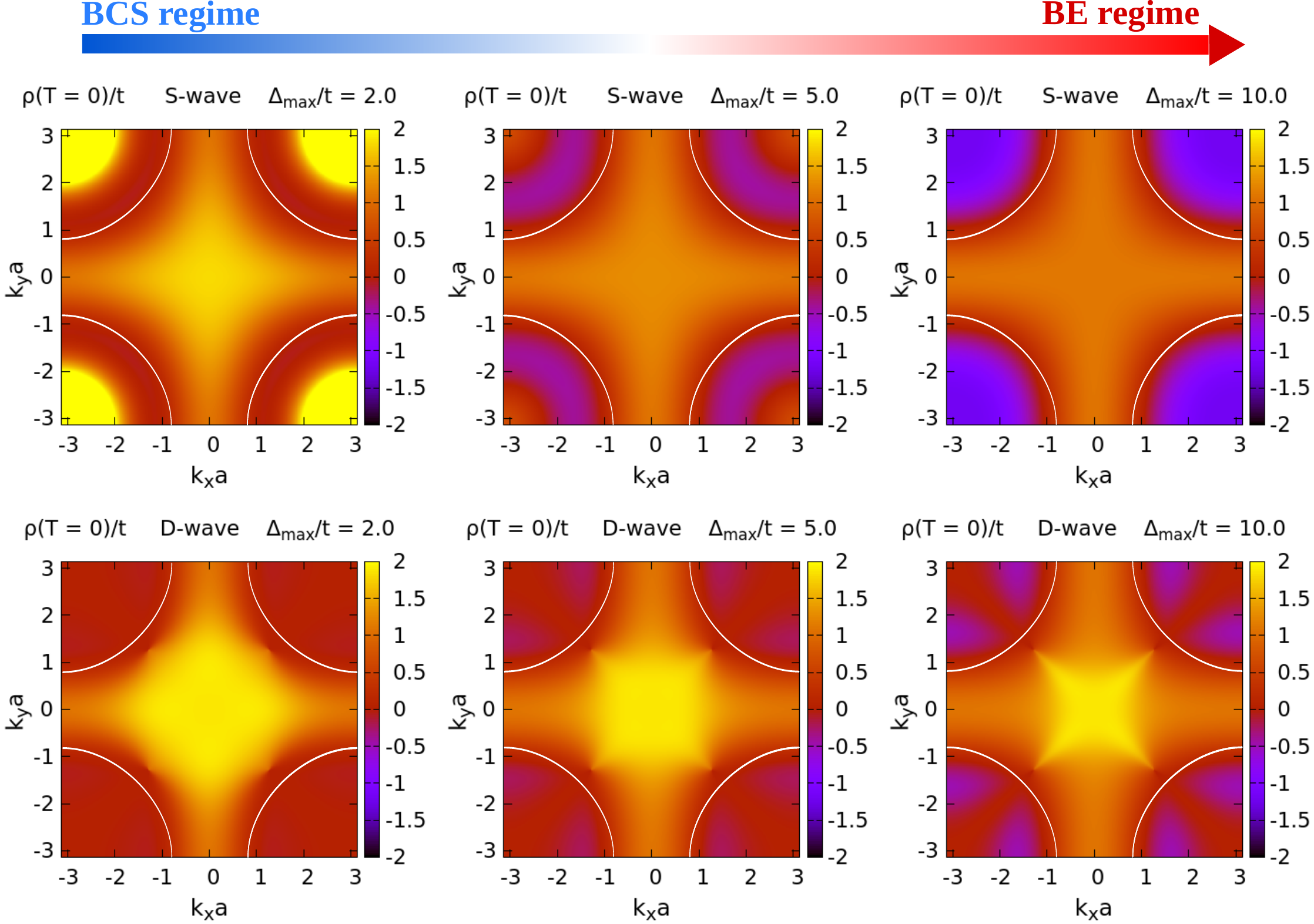}
 \caption{Phase stiffness at zero temperature for the $s$-wave and $d$-wave channels is plotted over the BZ and reported for different superconducting gaps. The chemical potential $\mu$ is fixed at VHs = -4t$^{\prime}$. White line indicates the Fermi level.}\label{fig.8}
\end{figure*}

For the given energy dispersion with negligible next-nearest-neighbor hopping, the BKT transition temperature for both $s$- and $d$-wave pairings is plotted as a function of the mean-field temperature shown in Fig.~\ref{fig.5}. Interestingly, in the BCS regime, the critical temperatures (T$_{\text{c}}$, T$_{\text{BKT}}$) for both pairing symmetries are nearly identical, despite the significantly different pairing strengths, as shown in the inset of Fig.~\ref{fig.5}. When the mean-field T$_{c}$ approaches approximately 10$\%$ of $t$, the BKT transition temperatures associated with $s$-wave and $d$-wave pairing symmetries begin to diverge appreciably from the mean-field T$_{c}$, marking the onset of the pseudogap regime. Within this crossover region, the $s$-wave BKT temperature remains marginally higher than the $d$-wave counterpart. Upon entering the BE regime, the pseudogap magnitude increases significantly for $s$-wave pairing, whereas it remains comparatively suppressed for the $d$-wave symmetry. In this limit, the $d$-wave BKT transition temperature becomes approximately four times larger than that of the $s$-wave. This behavior directly results from shrinking of the pseudogap, which typically arises from very low superconducting fluctuations. In the BE regime, the mean-field critical temperature is high; as a result, one expects strong superconducting fluctuations in the form of pre-formed Cooper pairs, which destroy long-range order and suppress superconductivity. Consequently, the $s$-wave symmetry exhibits a very low BKT transition temperature. On the other hand, the $d$-wave symmetry, with the presence of nodal quasiparticles, screens these fluctuations and enhances the BKT transition temperature. The zero-temperature phase stiffness, with and without the inclusion of the next-nearest-neighbor hopping parameter, is plotted as a function of the mean-field critical temperature in Fig.~\ref{fig.6}. The inset shows the band-dependent DOS as a function of energy, expressed in units of the hopping parameter. In the case of negligible next-nearest-neighbor hopping, a pronounced VHs appears in the DOS, which significantly suppresses the phase stiffness for $s$-wave pairing, while its effect on the $d$-wave symmetry remains minimal. When next-nearest-neighbor hopping is included, the VHs is broadened and shifted, thereby mitigating its impact.

We extended our investigation of pairing symmetries as a function of carrier density by excluding next-nearest-neighbor hopping, as shown in Fig.~\ref{fig.7}. At low carrier densities a$^{2}$n = 0.15 and 0.25, the parabolic region of the energy band is accessed, and there is no dominance of the $d$-wave channel even in the strong-pairing regime. Nonetheless, just above the filling of a$^{2}$n = 0.5, the pairing of the $d$-wave channel begins to dominate over the $s$-wave one when T$_{\text{c}}\approx t$. This transition arises because, beyond quarter filling, the $s$-wave symmetry becomes increasingly sensitive to the VHs, leading to suppression of its BKT transition temperature \cite{sathish, Q.Gao}. This mechanism is not observed in the $d$-wave case due to the presence of distinct quasi-condensates that shield it from such effects.

\begin{figure*}[t]
\centering
\includegraphics [width = 0.75\textwidth]{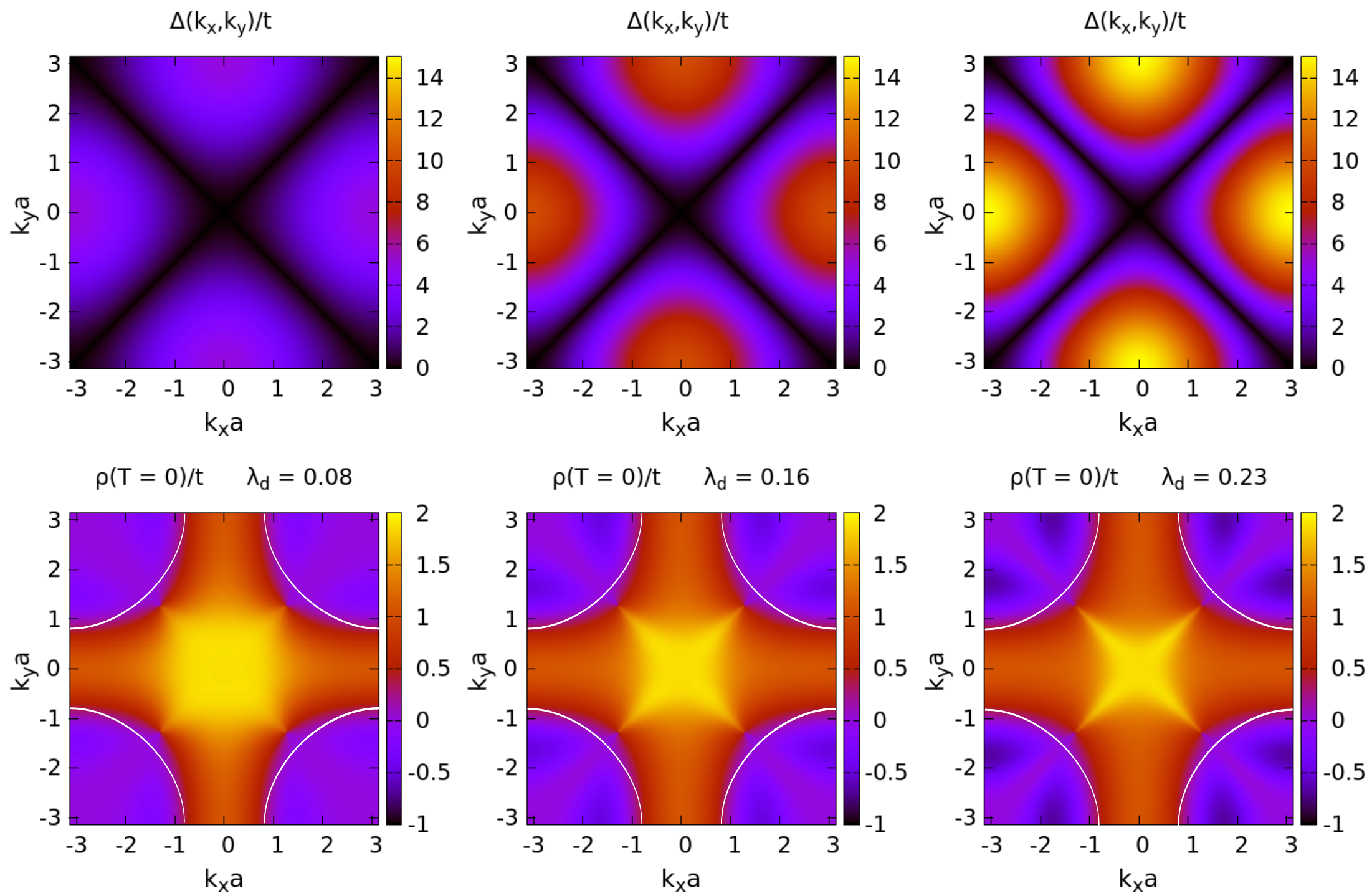}
 \caption{Top panels represent the intensity of the order parameter in units of the hopping strength $t$ in the $d$-wave pairing channel, over the entire BZ, for different pairing strengths. Bottom panels show the corresponding plots of phase stiffness in units of $t$. The chemical potential $\mu$ is fixed at VHs = -4t$^{\prime}$.}
    \label{fig.9}
\end{figure*}

Fig.~\ref{fig.8} compares the phase stiffness distribution across the BZ for $s$-wave and $d$-wave pairing at a representative superconducting gap. The phase stiffness was defined as $\rho_{s}$(T=0) = $\frac{\rho_{xx} + \rho_{yy}}{2}$. In both cases, the region near the $\Gamma$ point exhibits the highest stiffness due to the small effective mass and large Fermi velocity associated with the parabolic dispersion at the bottom of the band. For $d$-wave pairing in the strong-pairing regime, hot patches expand into areas previously dominated by cold patches, consistent with the pronounced inverse dependence of the superconducting gap on phase stiffness. Fig.~\ref{fig.9} shows the intensity maps of the phase stiffness and the $d$-wave order parameter over the BZ. In the weak-pairing limit, particularly near the nodal line, the phase stiffness remains relatively high and uniformly distributed. As the pairing strength increases, the superconducting gap grows predominantly in the hot patch regions, while the phase stiffness in the cold regions diminishes. This evolution illustrates that stronger pairing promotes the expansion of hot patches into cold regions, thereby reducing the maximum phase stiffness. Overall, these observations highlight the clear inverse relationship between the superconducting gap and the total phase stiffness in the strong-pairing regime.

\subsection{Two-component phase stiffness}
\begin{figure}[b]
\centering
\includegraphics[width= \linewidth]{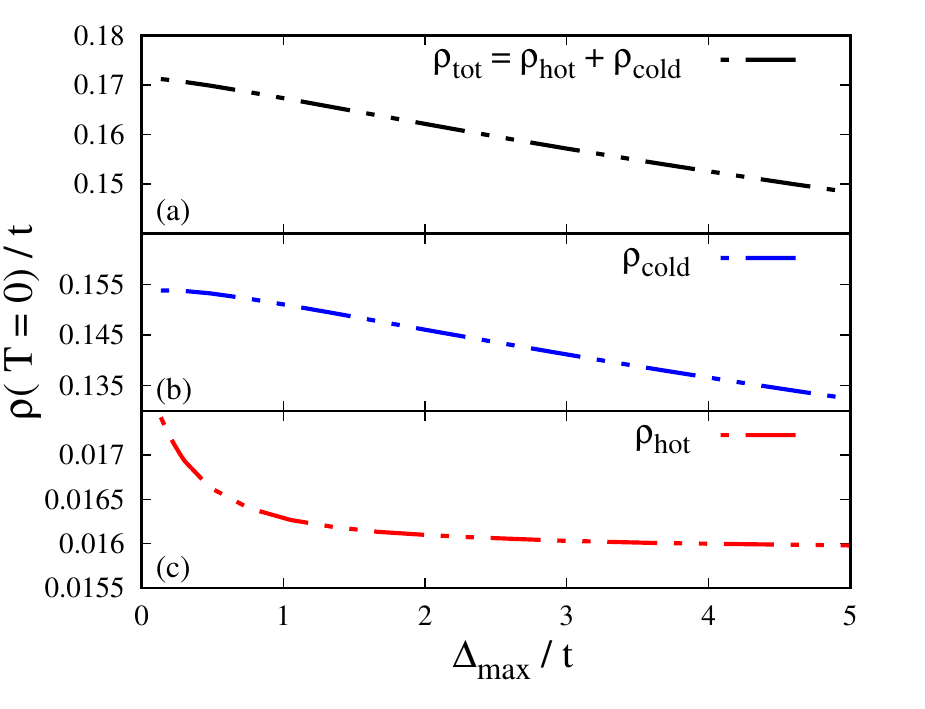}
 \caption{(a) Total phase stiffness as a function of the maximum superconducting gap in units of $t$. (b) Phase stiffness for the cold patches. (c) Phase stiffness for the hot patches.}
    \label{fig.10}
\end{figure}

In the spirit of the above discussed two-patch model, the results we obtained from the intensity plot of the stiffness motivated us to decouple the total stiffness into two components: one with cold patch stiffness where the stiffness is large and Cooper pairs are in the BCS regime, and another with hot patch stiffness where the stiffness is small and Cooper pairs are associated with the BE molecular regime, i.e.
\begin{equation}
    \rho_{tot} = \rho_{cold} + \rho_{hot}. 
\end{equation}
The phase stiffness within the cold patch encompasses the area surrounding the nodal line, whereas the phase stiffness associated with the hot patch is characterized by the presence of saddle points in the BZ and has a VHs in the DOS. In the strong-pairing regime, the phase stiffness associated with the hot patch experiences a more pronounced suppression when compared to the region corresponding to the cold patch. Fig.~\ref{fig.10} displays three panels: (a) the total phase stiffness at zero temperature, (b) phase stiffness for cold patches, and (c) phase stiffness for hot patches.
Interestingly, the phase stiffness of the cold patches in the vicinity of the nodal line exhibits a quadratic dependence on the superconducting gap. The phase stiffness of the cold patches can be approximated as
\begin{equation}\label{16}
    \rho_{cold}  \approx  \rho_{0} + \rho_{1}   - \rho_{2},
\end{equation}
where,
\begin{equation*}
    \rho_{0}  =  \frac{1}{4}\int \frac{d^{2}k}{(2\pi)^{2}} \frac{\partial^{2}\epsilon(\boldsymbol{k}) }{\partial k_{\alpha}^{2}} \left[ 1 - \frac{\xi_{\boldsymbol{k}} }{ \lvert \xi_{\boldsymbol{k}} \rvert  } \right],
\end{equation*}
\begin{equation*}
    \rho_{1} = \frac{1}{4} \int \frac{d^{2}k}{(2\pi)^{2}} \frac{\partial^{2} \epsilon(\boldsymbol{k})}{\partial k_{\alpha}^{2}} \left[ \frac{\Delta^{2}_{k}}{2 \lvert \xi_{\boldsymbol{k}} \rvert \xi_{\boldsymbol{k}}} \right],
\end{equation*}
\begin{equation*}
    \rho_{2} = \frac{1}{4} \int \frac{d^{2}k}{(2\pi)^{2}} \frac{\partial^{2} \epsilon(\boldsymbol{k})}{\partial k_{\alpha}^{2}} \left[ \frac{3 \Delta^{4}_{k}}{8 \lvert \xi_{\boldsymbol{k}} \rvert \xi^{2}_{\boldsymbol{k}}} \right].
\end{equation*}

\begin{figure}[b]
\centering
\includegraphics[width= \linewidth ]{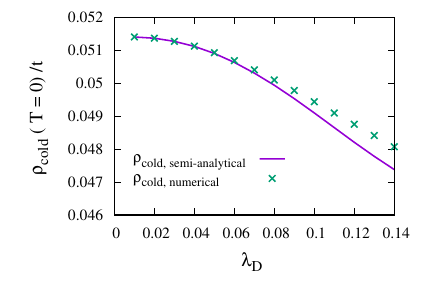}
 \caption{Phase stiffness of the cold patch at $T$ = 0 as a function of the pairing strength ($\lambda_{D}$).}
\label{fig.11}
\end{figure} 
The $\rho_{0}$ contribution originates from the nodal line of the $d$-wave order parameter, where the superconducting gap is zero. Fig.~\ref{fig.11} shows the evolution of phase stiffness at zero temperature with the strength of the $d$-wave pairing. In the realm of weak-pairing, the semi-analytical results from Eqn.~(\ref{16}) match the numerical results. For very weak-pairing strength, the superconducting gap is small, rendering it practically negligible and resulting in saturated phase stiffness. However, upon entering a strong-pairing regime, a discernible deviation emerges between the numerical and semi-analytical results. In close proximity to the nodal lines, the phase stiffness is directly proportional to the superconducting gap, which exhibits a quadratic nature. In the hot region however, phase stiffness is inversely proportional to the superconducting gap. In Fig.~\ref{fig.1}, the behavior of the phase stiffness is shown in two different regions of the BZ. Recent experiments in monolayer Bi2212 indicate that the critical temperature values are not significantly different from those observed in 3D bulk counterparts in the optimal doping regime~\cite{mono2,zhang_2023}. Here, we fit our data with the experimental observation; in order to achieve a more realistic description of monolayer cuprates, we opt for the hopping parameter $t$ = 0.15~eV. We fixed the chemical potential at the VHs, which corresponds to the optimal doping regime. Our obtained T$_{\text{BKT}}$ value of 90.17~K, and the onset-pairing critical temperature T$_{\text{c}}$ of 114.3~K. The experimental BKT temperature is 89.1~K \cite{zhang_2023}. Additionally, the maximum superconducting gap observed in the experiment is 47.3~meV, whereas our results yield 41.6~meV. The corresponding effective pairing strength outside the VHs region is 0.031. These results match reasonably well with experimental values and we predict the existence of a pseudogap in the range of temperatures 25 K above the BKT transition, which is very close to the measured pseudogap temperature range for monolayer Bi2212 at optimal doping. Although high-T$_{\textbf{c}}$ cuprates predominantly reside within the crossover regime, close to the BCS weak-pairing regime, the present analysis establishes a versatile framework for investigating prospective unconventional high-T$_{\textbf{c}}$  superconducting materials characterized by $d$-wave symmetry in the 2D limit, irrespective of their multiband or multigap nature. In particular, this approach offers valuable insights into the BKT physics including phase stiffness, the BKT transition temperature, and the pseudogap of such materials.

\section{Conclusions}
Motivated by the search for high BKT transition temperatures in strongly coupled 2D superconductors, in this work we discuss how $d$-wave symmetry of the superconducting gap can be advantageous compared to $s$-wave superconductors. For $s$-wave gaps, the phase stiffness is substantially suppressed close to the Van Hove singularity (VHs), leading to low BKT transition temperatures \cite{sathish}. We thus presented a comparative analysis of $s$-wave and $d$-wave symmetries of the superconducting gap to explore their impact on the phase stiffness and the BKT transition temperature in strongly coupled 2D superconductors. Our study revealed that in the strong-pairing regime, the phase stiffness and the BKT transition temperature of the $d$-wave superconductor exceed those of the $s$-wave one for an equivalent maximum superconducting gap or the same mean-field pairing temperature scale. This enhancement is attributed to the anisotropic wave-vector-dependent superconducting gap. We have highlighted the phase stiffness intensity in the BZ at zero temperature, revealing the coexistence of two distinct condensate components. The overall phase stiffness is composed of two contributions: one with high phase stiffness around the nodal lines, and another with low phase stiffness around the saddle points of the BZ (maximum gap). The coexistence of the BCS and BE-like pairing regimes within a single band, induced with an anisotropic pairing potential, stabilizes a high BKT transition temperature by suppressing the phase fluctuations in the 2D limit. Within the BCS–BE crossover regime, we comprehensively discussed the universal behavior, the evolution of the pseudogap, and the interplay between the BKT transition temperatures of the $s$- and $d$-wave gap parameters relative to the mean-field T$_{c}$. Experimental results from few- or single-unit cell cuprates, which exhibit high critical temperatures in the 2D regime with $d$-wave pairing symmetry, could be understood by the physics of a two-component phase stiffness model, as here described. To accurately capture the realistic BKT transition temperature of 2D cuprates, we fitted our model data and presented the results for critical temperatures and the superconducting gap. This effective two-component analysis demonstrates a crossover behavior of the phase stiffness relative to the wave-vector dependence of the superconducting gap from the nodal lines toward the M points of the BZ. The emergence of this dual-component phase stiffness in $d$-wave superconductors provides a valuable framework for understanding the physics of high-T$_{\text{c}}$ 2D superconductors, highlighting the importance of the wave-vector-anisotropic pairing and of strongly wave-vector-dependent superconducting gaps, resulting in amplification of the total phase stiffness of the superconducting state. Our approach to analyzing the $d$-wave gap in 2D superconductors is broadly applicable across diverse pairing regimes, carrier densities, and systems exhibiting VHs, including higher-order singularities, and are not confined to single-band or one-gap $d$-wave superconducting systems; instead they are broadly applicable to multi-band or multi-gap superconducting systems, provided that at least one $d$-wave partial condensate is present. 

\section*{Acknowledgements}
We acknowledge Jonas Bekaert and Giovanni Midei for useful discussions and critical reading of the manuscript. This work has been supported by the PNRR MUR project PE0000023-NQSTI and by the Research Foundation-Flanders (FWO-Vl).

\bibliography{ref}
\end{document}